\documentclass[12pt]{article}
\usepackage{graphicx}
\usepackage{slashed}
\DeclareGraphicsExtensions{.pdf}
\usepackage{float} 
\usepackage{amsmath}
\usepackage{amsfonts}   
\usepackage{amssymb}

\begin{document}
\date{}
\title{{\bf{\Large Holography for anisotropic branes with hyperscaling violation}}}
\author{
 {\bf {\normalsize Dibakar Roychowdhury}$
$\thanks{E-mail:  dibakarphys@gmail.com, dibakarr@iitk.ac.in}}\\
 {\normalsize  Indian Institute of Technology, Department of Physics,}\\
  {\normalsize Kanpur 208016, Uttar Pradesh, India}
}

\maketitle
\begin{abstract}
In this paper, based on the principles of Gauge/gavity duality, we explore the field theory description of certain special class of strongly coupled hyperscaling violating QFTs in the presence of scalar deformations near the \textit{effective} dynamical scale ($ r_F $) of the theory. In the language of the AdS/CFT duality, the scalar deformations of the above type could be thought of as being sourced due to some massless scalar excitation in the bulk which explicitly break the  $ SO(2) $ rotational invariance along the spatial directions of the brane. As a consequence of these deformations, it turns out that when we probe such QFTs in terms of its non-local observable like, the entanglement entropy as well as the Wilson operator they indeed receive  finite contributions near the effective dynamical scale ($ r_F $) of the theory.
\end{abstract}

\section{Overview and Motivation}
For the past couple of years, the holographic theories with hyperscaling violation have received renewed attention due to their remarkable connection to that with the known condensed matter systems \cite{Dong:2012se}-\cite{Narayan:2012hk}, specially the physics near the Fermi surface \cite{Ogawa:2011bz}-\cite{Alishahiha:2015goa}. In the framework of the AdS/CFT duality, the hyperscaling violating QFTs could be geometrically realized in their dual description as \cite{Alishahiha:2012qu}, 
\begin{eqnarray}
ds^{2}=r^{\frac{-2 \theta}{d}}\left( - r^{2z}dt^{2}+\frac{dr^{2}}{r^{2}}+r^{2}dx_{i}^{2}\right),~~(i=1,...,d)
\label{I1}
\end{eqnarray}
where, $ z $ is the dynamical critical exponent and $ \theta $ is the so called hyperscaling violating exponent. The geometry of the above type (\ref{I1}), could be consistently realized in a bottom up approach that includes Einstein-Maxwell-Dilaton (EMD) gravity in the bulk \cite{Alishahiha:2012qu}.

The most notable fact about the above metric (\ref{I1}) is the existence of the obvious scaling symmetries of the type \cite{Alishahiha:2012qu},
\begin{eqnarray}
t \rightarrow \lambda^{z}t,~~r\rightarrow r/\lambda ,~~x_{i} \rightarrow \lambda x_i,~~ds_{d+2} \rightarrow \lambda^{\theta / d}ds_{d+2}.
\label{ee1}
\end{eqnarray}

Now, before we proceed further, a few important remarks are in order. First of all, one should note that the geometry of the above type (\ref{I1}) appears as an \textit{effective} holographic description (of some strongly coupled QFT) which is valid below certain specific energy scale of the theory. More specifically, hyperscaling violating QFTs could be thought of as an \textit{effective} theory that appears as a low energy description of some UV complete field theory. It might so happen that the QFT that originally had a UV fixed point, now develops an additional dimensionful scale\footnote{For a system consisting of fermions, this energy scale could be thought of as the radius of the Fermi sphere in the momentum space.} $ r_F (\gg r_{UV})$ during its RG flow towards lower energies that might not get decoupled in the deep infra red regime of the theory \cite{Dong:2012se}. The appearance of such an additional dynamical scale ($ r_F $) is often identified as the key source of hyperscaling violation at lower energies. Under such circumstances, the space time metric in the dual gravitational counterpart scales as, $ ds^{2}\sim \frac{L^{2}_{AdS}}{r_{F}^{2\theta/d}} $, where, $ L_{AdS} $ is the so called AdS length scale that we set equal to one in the subsequent analysis of this paper.  

In our analysis, however, we consider $ U(1) $ charged black brane configurations \cite{Alishahiha:2012qu} that asymptotically approach the above geometry (\ref{I1}). In other words, in our analysis we put our \textit{boundary} QFT at an effective energy scale, $ E_{F}\sim r_F^{-1}(\ll E_{UV}) $. It is the neighbourhood of this energy scale where we would confine ourselves throughout the entire analysis. In particular, our focus would be to explore the behaviour of a few non local observable of the theory when the overall theory is deformed due to some scalar perturbations near its effective energy scale ($ \sim E_{F} $). The overall energy range for our analysis is therefore turned out to be, $ r^{-1}_H < E \leq r_F^{-1} $, where $ r_H $ denotes the location of the horizon of the black brane such that $ r_F \ll r_H $. Therefore, $ r=r_F $, literally acts as an \textit{effective} UV cut off (namely, the \textit{boundary}) for our analysis. There might exists some true UV fixed point for $ r \ll r_F $ (in case of asymptotic AdS spaces), but that is something which we do not care at this moment. 

In several earlier analysis \cite{Dong:2012se}-\cite{Shaghoulian:2011aa}, it has been confirmed that the hyperscaling violating QFTs those are dual to (\ref{I1}), might reveal important phase structure when probed through non local observable like the entanglement entropy of certain subregion of the system. In particular, it has been confirmed that for $ \theta =d-1 $, one indeed encounters the logarithmic violation of the area law which clearly indicates the existence of hidden Fermi surfaces for the boundary theory \cite{Dong:2012se}-\cite{Shaghoulian:2011aa}. In \cite{Dong:2012se}, the authors have further explored the entanglement entropy for hyperscaling violating QFTs at finite temperature and computed thermal corrections both in the small as well as in the large temperature regimes. It turns out that at finite temperature, the entanglement entropy receives non trivial contributions due to thermal effects present in the system.

Keeping the spirit of the above discussions, the purpose of the present article is therefore to extend the above analysis in the presence of scalar deformations near the effective UV scale ($ E \sim E_F $) of the theory. More precisely, we would like to understand how the non local observable of the boundary theory namely, the entanglement entropy as well as the Wilson operator capture the signature of the presence of such perturbations in the original theory. In the language of the AdS/CFT duality, such deformations at the boundary are sourced due to some massless scalar excitation in the bulk which explicitly break the $ SO(2) $ rotational invariance along the spatial directions of the brane \cite{Mateos:2011tv}-\cite{Iizuka:2012wt} which thereby induce spatial anisotropy in the dual gravitational counterpart of the theory \footnote{Holographic models with spatial anisotropy have recently gained renewed attention due to its remarkable implications on various properties of the boundary plasma in particular, the hydrodynamic transport coefficients \cite{Rebhan:2011vd} as well as the thermalization \cite{Ecker:2015kna} at strong coupling.}. 

We now briefly outline the the key features of our analysis. In order to explore the effects of anisotropy on non local observable at the boundary, we first determine the background space-time perturbatively in the anisotropy parameter and we confine ourselves upto \textit{leading} order in the anisotropy. Remember that in our analysis, we are interested to explore any such (perturbative) effects of anisotropy close to the effective UV scale ($E \sim E_F $) of the boundary QFT. All these are done in Section 2.

In Section 3, we explicitly compute the holographic entanglement entropy (HEE) \cite{Ryu:2006bv}-\cite{Ryu:2006ef} confining ourselves to that of the low temperature regime of the theory namely, $ r\ll r_H $. In other words, we perform our bulk calculations at a scale, $ r\sim r_F $ which is therefore in a region close to the so called boundary of our space time. In our analysis, we focus on the special case namely \footnote{One of the primary motivations of our analysis turns out to be the study of anisotropic structure of Fermi surfaces for the electrons under a holographic set up. As it is quite well known that due to the presence of the crystal structure of the background lattices, the so called Fermi surface for electrons is not quite isotropic in the momentum space \cite{Iizuka:2012wt}.  Under such circumstances the holographic understanding of these anisotropic systems might shed new light into the subject. }, $ \theta =d-1 $. It turns out that under such circumstances, apart from having the usual logarithmic divergences \cite{Dong:2012se}-\cite{Shaghoulian:2011aa}, the HEE also receives finite contributions due to the presence of the anisotropy in the dual gravitational counterpart. As a matter of fact, it turns out that the anisotropic contributions to the HEE could be divided precisely into two pieces, one that do not associate with any temperature effects and the other, that yield finite temperature corrections to the entanglement entropy of the system.

In Section 4, we explore similar effects for another class of non local observable of the theory namely, the Wilson operator \cite{Maldacena:1998im}. It turns out that, the anisotropy produces a screening effect between the quark anti-quark ($ \mathfrak{q}\bar{\mathfrak{q}} $) pair at the boundary which in turn reduces the potential energy of the system. 

Finally, we conclude in Section 5.\\
$ \bullet $ \textbf{Note added:} Before we proceed further, let us first discuss some basic ingredients as well as qualitative features of the dual QFT under consideration. The dual QFT for the present example  could be thought of as being described in terms of \textit{de-confined} compressible states of some gapless (\textit{quark} like) excitation \cite{Huijse:2011ef}. In other words, the gauge fields in the theory do not confine which in turn result in so called gauge charged fermionic (quark like) excitation which are different than the excitation of the  weakly coupled version of the theory. These fractionalized fermionic excitation are supposed to be distributed over the so called \textit{hidden} Fermi surfaces as in the case of a non Fermi liquid \cite{Huijse:2011hp}-\cite{Sachdev:2010um}. Under a holographic set up, the existence of such hidden quark (Fermi) surfaces could be realized by computing the entanglement entropy which exhibits the so called logarithmic violation of the area law \cite{Ogawa:2011bz}. 

We now talk about the symmetries of the dual QFT under consideration. From the explicit construction in the bulk, it is indeed quite evident that the underlying symmetry group of the dual QFT near its UV (which is precisely the effective dynamical scale of the theory) fixed point is precisely that of the Killing isometry group generated by the Killing vector fields associated with the \textit{asymptotic} space time (\ref{I1}) which encompasses the so called generators of the rotation group associated with spatial coordinates ($ \textbf{x} $) of the brane and also contains dilataion generators corresponding to the anisotropic scaling symmetries of the above form (\ref{ee1}).
However, in the subsequent analysis, as we deform the theory by turning on  relevant scalar deformation of the form,
\begin{eqnarray}
\delta S_{QFT}\sim \int \Psi (x) Tr F \wedge F
\end{eqnarray}
the theory (RG) flows from its UV fixed point towards IR. Here, the operator $ \Psi (x) $ could be thought of as being sourced due to some scalar excitation in the bulk which linearly depends on one of the spatial coordinates of the  brane and as a consequence of this clearly breaks the rotational symmetry as well as the parity invariance associated with Fermi surfaces. However, after performing the integration by parts, we note that,
\begin{eqnarray}
\delta S_{QFT}\sim \int  dx Tr A \wedge F
\end{eqnarray}
which clearly exhibits the so called translation invariance of the theory.

\section{Anisotropic branes}
\subsection{The bulk set up}
The purpose of the present discussion is to cook up an anisotropic background which is dual to certain special class of (strongly coupled) \textit{effective} QFTs that do not preserve hyperscaling near its effective dynamical scale ($ r_F $). The dual theory that we consider in the bulk could be described in terms of Einstein-Maxwell Dilaton gravity with two abelian gauge fields. The first abelian one form together with the dilaton ($ \varphi $) is required in order to generate the anisotropic scaling in the solutions. Whereas, on the other hand, the black brane is supposed to be charged under the second $ U(1) $ field. The resulting action in ($ 3+1 $) dimensions could be formally expressed as \cite{Alishahiha:2012qu}, 
\begin{eqnarray}
S_{H}=-\frac{1}{16 \pi G}\int d^{4}x \left(R-\frac{1}{2}(\partial \varphi)^{2}+\mathsf{V}(\varphi)-\frac{1}{4}\sum_{i=1}^{2}\exp (\lambda_{i}\varphi) \mathsf{F}_{i}^{2}\right)
\label{E1} 
\end{eqnarray} 
where, $ \varphi $ is the scalar field and $ \mathsf{V}(\varphi)=\mathsf{V}_{0}\exp(\gamma \varphi) $, is the interaction potential for the scalar filed configuration together with $ \lambda_{1} $, $ \lambda_{2} $ and $ \mathsf{V}_{0} $ as the free parameters of the model \cite{Alishahiha:2012qu}. This interaction potential plays a significant role in order to trigger a non trivial hyperscaling violation for the boundary field theory.

The dynamics that immediately follows from the above action (\ref{E1}) could be formally expressed as \cite{Alishahiha:2012qu},
\begin{eqnarray}
R_{ab}&=&\frac{1}{2}\left(\partial_{a}\varphi \partial_{b}\varphi - \mathsf{V}(\varphi)g_{ab} \right)+\frac{1}{2}\sum_{i=1}^{2}\exp(\lambda_{i}\varphi) \left( \mathsf{F}^{c}\ _{ia}\mathsf{F}_{icb}-\frac{g_{ab}}{4}\mathsf{F}_{i}^{2}\right)=\mathsf{T}_{ab}\nonumber\\
\nabla^{2}\varphi &=&-\frac{\partial \mathsf{V}}{\partial \varphi}+\frac{1}{4}\sum_{i=1}^{2}\lambda_{i}\exp (\lambda_{i}\varphi)\mathsf{F}_{i}^{2}
\label{E2}
\end{eqnarray}
and,
\begin{eqnarray}
\nabla_{a}(\exp(\lambda_{i}\varphi)\mathsf{F}^{ab}_{i})=0. 
\label{E3}
\end{eqnarray}

The corresponding hyperscaling violating black brane configuration turns out to be\footnote{As mentioned earlier, in our analysis we have set the AdS length scale equal to one.} \cite{Alishahiha:2012qu},
\begin{eqnarray}
ds^{(0)2}&=&\left(\frac{r_H}{u} \right)^{-\theta} \left(-\left( \frac{r_H}{u}\right)^{2z}f^{(0)}(u)dt^{2}+\frac{du^{2}}{u^{2}f^{(0)}(u)}+\left(\frac{r_H}{u} \right)^{2}(dx^{2}+dy^{2})\right)\nonumber\\
f^{(0)}(u)&=&1-\mathsf{M}\left(\frac{u}{r_H} \right)^{z+2-\theta} + \mathsf{Q}^{2}\left(\frac{u}{r_H} \right)^{2(z+1-\theta)}\nonumber\\
\mathsf{F}_{1ut}^{(0)}&=&-\sqrt{2(z-1)(z+2-\theta)}\exp \left( \frac{(2-\frac{\theta}{2})\varphi_{0}}{\sqrt{2(2-\theta)(z-1-\frac{\theta}{2})}}\right)\frac{r_H^{z-\theta +2}}{u^{z-\theta +3}}\nonumber\\
\mathsf{F}_{2ut}^{(0)}&=&-\mathsf{Q}\sqrt{2(2-\theta)(z-\theta)}\exp \left(\sqrt{\frac{z-1-\frac{\theta}{2}}{2(2-\theta)}} \varphi_{0}\right)\frac{u^{z-\theta -1}}{r_H^{z-\theta}}\nonumber\\
\exp \varphi &=& \exp \varphi_{0}\left( \frac{r_H}{u}\right)^{\sqrt{2(2-\theta)(z-1-\theta /2)}}
\label{E4}  
\end{eqnarray}
Here, $ \mathsf{M} $ is the mass and $ \mathsf{Q} $ is the abelian charge of the black brane \cite{Alishahiha:2012qu}. Note that, in the above we have expressed the hyperscaling violating black brane configuration (\ref{E4}) in terms of a modified radial variable $ u(=r_H/r) $ such that the horizon of the black brane is placed at $ u=1 $, while on the other hand, the boundary of the space time is located in the limit, $ u \rightarrow 0 $. 
Furthermore, from the above solution (\ref{E4}) we note that the $ SO(2) $ rotational invariance is indeed preserved along the spatial directions of the brane which thereby reflects the isotropy. In the first part of our analysis, the non trivial task therefore would be to search for the corresponding anisotropic description in the presence of the hyperscaling violation such that the anisotropy vanishes near the effective UV scale ($ \sim E_F $) of the boundary theory.

\subsection{The solution}
Before we formally elaborate the details regarding our anisotropic construction, it is customary to mention that the anisotropy that we consider in this paper could be thought of as perturbations over the original (unperturbed) background (\ref{E4}) where we restrict ourselves only upto leading order in the anisotropic fluctuations. 

In order to proceed further, we choose the following metric ansatz for the $ U(1) $ charged hyperscaling violating black brane configuration in ($ 3+1 $) dimensions namely \cite{Iizuka:2012wt},
\begin{eqnarray}
ds^{2}=\left(\frac{r_H}{u} \right)^{-\theta} \left(-\left( \frac{r_H}{u}\right)^{2z}f(u)dt^{2}+\frac{du^{2}}{u^{2}f(u)}+e^{\mathcal{A}(u)+\mathcal{B}(u)}dx^{2}+e^{\mathcal{A}(u)-\mathcal{B}(u)}dy^{2}\right).
\label{E5}
\end{eqnarray} 

From the above black brane configuration (\ref{E5}), it is indeed evident that the earlier $ SO(2) $ rotational invariance is now broken due to the presence of the function $ \mathcal{B}(u) $. However, the homogeneity is still retained at a fixed radial hypersurface $ u=u_{c} $, whose generators are the generators of the Killing isometries namely, $ \partial_{x} $ and $ \partial_{y} $. 

The corresponding Einstein's equations (\ref{E2}) turn out to be,
\begin{eqnarray*}
\frac{1}{2} r_H^{2 z-\theta } \left(\frac{r_H}{u}\right)^{\theta } u^{-2 z+\theta } f \left((2 z-\theta ) f \left(z-\theta -u \mathcal{A}'\right)+u \left(\left(1-3 z+2 \theta +u \mathcal{A}'\right) f'+u f''\right)\right)=\mathsf{T}_{tt}
\end{eqnarray*}
\begin{eqnarray*}
f \left(z (2 z-\theta )+u \left((2+\theta ) \mathcal{A}'+u \mathcal{A}'^2+u \left(\mathcal{B}'^2+2 \mathcal{A}''\right)\right)\right)+u \left(\left(1-3 z+2 \theta +u \mathcal{A}'\right) f'+u f''\right)\nonumber\\
=-2u^{2}f\mathsf{T}_{uu}
\end{eqnarray*}
\begin{eqnarray*}
u \left(\theta +u \left(\mathcal{A}'+\mathcal{B}'\right)\right) f'+f \left(\theta  (-z+\theta )+u \left(u \mathcal{A}'^2+(1-z+\theta ) \mathcal{B}'+\mathcal{A}' \left(1-z+2 \theta +u \mathcal{B}'\right)+u \left(\mathcal{A}''+\mathcal{B}''\right)\right)\right)\nonumber\\
=-2e^{-(\mathcal{A}+\mathcal{B})}\mathsf{T}_{xx}
\end{eqnarray*}
\begin{eqnarray}
u \left(\theta +u \mathcal{A}'-u \mathcal{B}'\right) f'-f \left((z-\theta ) \theta +u \left(-u \mathcal{A}'^2+(1-z+\theta ) \mathcal{B}'+\mathcal{A}' \left(-1+z-2 \theta +u \mathcal{B}'\right)-u \left(\mathcal{A}''-\mathcal{B}''\right)\right)\right)\nonumber\\
=-2e^{-\mathcal{A}+\mathcal{B}}\mathsf{T}_{yy}\nonumber\\
\end{eqnarray}
which could be further re-arranged as,
\begin{eqnarray*}
\mathcal{B}''+\mathcal{B}'\left(\mathcal{A}'+\frac{1-z+\theta}{u}+\frac{f'}{f} \right)&=&-\frac{e^{-\mathcal{A}}}{u^{2}f}(e^{-\mathcal{B}}\mathsf{T}_{xx}-e^{\mathcal{B}}\mathsf{T}_{yy}),\nonumber\\
\mathcal{A}''+\mathcal{A}'^{2}+\mathcal{A}'\left(\frac{f'}{f}+\frac{1-z+2\theta}{u} \right)+\frac{\theta(\theta -z)}{u^{2}}+\frac{\theta f'}{u f}&=&-\frac{e^{-\mathcal{A}}}{u^{2}f}(e^{-\mathcal{B}}\mathsf{T}_{xx}+e^{\mathcal{B}}\mathsf{T}_{yy}), 
\end{eqnarray*}
\begin{eqnarray}
f'' +\frac{f'}{u}(1-3z+2\theta+u\mathcal{A}')+\frac{f(2z-\theta)(f(z-\theta)+z)}{u^{2}(f+1)}-\frac{f(\mathcal{A}'^{2}-\mathcal{B}'^{2})}{(f+1)}\nonumber\\
+\frac{\mathcal{A}'}{u^{2}(f+1)}\left(uf(2+\theta)-uf^{2}(2z-\theta)-2u^{2}f\left(\frac{f'}{f}+\frac{1-z+2\theta}{u} \right)  \right)\nonumber\\
-\frac{2(\theta f (\theta -z)+u \theta f')}{u^{2}(f+1)}- \frac{2e^{-\mathcal{A}}}{u^{2}(f+1)}(e^{-\mathcal{B}}\mathsf{T}_{xx}+e^{\mathcal{B}}\mathsf{T}_{yy})=\frac{2f}{(f+1)}\left( \frac{u^{2(z-1)}}{f r_{H}^{2z}}T_{tt}-T_{uu}\right). 
\label{E11}
\end{eqnarray}

In order to proceed further, we choose the following ansatz for the $ U(1) $ gauge fields namely,
\begin{eqnarray}
\mathsf{A}_{i m}dx^{m}=\mathsf{A}_{i t}(u)dt.
\end{eqnarray}
If we further demand that $ \varphi =\varphi (u) $, then the R.H.S. of the first equation of (\ref{E11}) vanishes identically,
\begin{eqnarray}
e^{-\mathcal{A}}(e^{-\mathcal{B}}\mathsf{T}_{xx}-e^{\mathcal{B}}\mathsf{T}_{yy})=0.
\end{eqnarray}

As a consequence of this, from the first equation in (\ref{E11}) we note that,
\begin{eqnarray}
\mathcal{B}'(u)= \mathit{C} e^{-\int_{1}^{u}\left(\mathcal{A}'(u')+\frac{1-z+\theta}{u'}+\frac{f'(u')}{f(u')} \right)du'}. 
\label{E13}
\end{eqnarray}

Note that, the integrand above in (\ref{E13}) clearly blows up in the near horizon ($ u'=1 $) limit due to the vanishing of the function $ f(u) $ there.  This observation has got some deeper consequences. It essentially implies that if one starts with the original isotropic configuration (\ref{E4}) and try to perturb it by turning on an anisotropy of the form $ \mathcal{B}(u) $ then this automatically leads to inconsistencies in the solution as the radial flow of the function $ \mathcal{B}(u) $ turns out to be ill defined which implies that the metric is not regular everywhere in the space time. The consistency in the solutions could be retried back iff we set $\mathit{C}=0$, which thereby implies isotropy \cite{Iizuka:2012wt}. 

 As we shall shortly see that in order to retain the anisotropy ($ \mathcal{B}\neq 0 $), one must have a non zero contribution on the R.H.S. of the first equation in (\ref{E11}) namely,
\begin{eqnarray}
T^{x}\ _{x}\neq T^{y}\ _{y}.
\end{eqnarray}

In order to generate such anisotropies we deform our original action (\ref{E1}) by a scalar perturbation of the following form \cite{Iizuka:2012wt},
\begin{eqnarray}
S \rightarrow S_{H}-\frac{1}{2}\int (\partial \Phi)^{2}
\end{eqnarray}
where, 
\begin{eqnarray}
\Phi = \Lambda x
\label{e16}
\end{eqnarray}
could be thought of as the source to the anisotropy\footnote{In the language of the dual field theory, the deformation of the above type (\ref{e16}) could be thought of as deformations of the boundary theory where we deform our boundary action by turning on the corresponding scalar operator dual to $ \Phi (x) $ in the bulk \cite{Mateos:2011tv}.} that ultimately leads to the desired anisotropy of the following form namely,
\begin{eqnarray}
T^{x}\ _{x}- T^{y}\ _{y}=\frac{\Lambda^{2}}{2}\left(\frac{r_H}{u} \right)^{\theta}e^{-(\mathcal{A}+\mathcal{B})}.
\label{e17} 
\end{eqnarray}
From (\ref{e17}), it is indeed evident that the effect of anisotropy appears with the quadratic power in $ \Lambda $ which thereby motivates us to solve the background (\ref{E5}) perturbatively upto quadratic order in $ \Lambda $. 

Using (\ref{e16}) one can further simplify (\ref{E11}) as\footnote{The equation corresponding to $ f $ has been provided in Appendix I.},
\begin{eqnarray*}
\mathcal{B}''+\mathcal{B}'\left(\mathcal{A}'+\frac{1-z+\theta}{u}+\frac{f'}{f} \right)+\Lambda^{2}\frac{e^{-(\mathcal{A}+\mathcal{B})}}{2u^{2}f}=0,
\end{eqnarray*}
\begin{eqnarray*}
\mathcal{A}''+\mathcal{A}'^{2}+\mathcal{A}'\left(\frac{f'}{f}+\frac{1-z+2\theta}{u} \right)+\frac{\theta(\theta -z)}{u^{2}}+\frac{\theta f'}{u f}+\Lambda^{2}\frac{e^{-(\mathcal{A}+\mathcal{B})}}{2u^{2}f}-\frac{\mathsf{V}(\varphi)}{u^{2}f}\left(\frac{u}{r_H} \right)^{\theta}\nonumber\\
-\frac{1}{4 u^{2}f}\left(\frac{u}{r_H} \right)^{\theta}\sum_{i}e^{\lambda_{i}\varphi}\mathsf{F}_{i}^{2} =0.
\label{E19}
\end{eqnarray*}

As a next step of our analysis, we solve the above set of equations (\ref{E19}) perturbatively in the (anisotropic) parameter $ \Lambda $ namely,
\begin{eqnarray}
\mathcal{A}&=& \mathcal{A}^{(0)}+\Lambda^{2}\mathcal{A}^{(1)}+ \mathcal{O}(\Lambda^{4})\nonumber\\
\mathcal{B}&=& \Lambda^{2}\mathcal{B}^{(1)}+\mathcal{O}(\Lambda^{4})\nonumber\\
f&=& f^{(0)}+\Lambda^{2} f^{(1)}+\mathcal{O}(\Lambda^{4})
\label{E20}
\end{eqnarray}
where we retain the effects of anisotropy only upto the quadratic order in $ \Lambda $. In order to match our solutions to that with the original unperturbed solution (\ref{E4}), the corresponding expression for the function $ \mathcal{A}^{(0)} $ turns out to be,
\begin{eqnarray}
\mathcal{A}^{(0)}(u)=2 \log \left(\frac{r_H}{u} \right). 
\end{eqnarray}

Before we proceed further, it is also customary to have a look at the constraints imposed by the anisotropy on the equations corresponding to the $ U(1) $ gauge field(s) as well as the the dilation ($ \varphi $). From (\ref{E3}), it is quite trivial to note down,
\begin{eqnarray}
e^{\lambda_{i}\varphi}\mathsf{F}_{iut}=-\frac{\mathsf{N}}{r_H^{2-z}u^{z-1}}(1-\Lambda^{2}\mathcal{A}^{(1)})+\mathcal{O}(\Lambda^{4})
\label{E21}
\end{eqnarray}
where, $\mathsf{N}$ is some constant such that we recover our original (unperturbed) results in the limit of the vanishing anisotropy.

Finally, substituting (\ref{E20}) into (\ref{E19}) we arrive at the following set of equations\footnote{The equation corresponding to $ f^{(1)} $ has been provided in Appendix II.},
\begin{eqnarray*}
\mathcal{B}''^{(1)}+\mathcal{B}'^{(1)}\left(\frac{\theta -z}{u}-\frac{1}{u}+\frac{f'^{(0)}}{f^{(0)}} \right)+\frac{1}{2r_{H}^{2}f^{(0)}}=0, 
\end{eqnarray*}
\begin{eqnarray*}
\mathcal{A}''^{(1)}+\mathcal{A}'^{(1)}\left(\frac{f'^{(0)}}{f^{(0)}}+\frac{-z+2\theta}{u}-\frac{3}{u} \right)+ \frac{f'^{(0)}}{f^{(0)}}\left(\frac{f'^{(1)}}{f'^{(0)}}-\frac{f^{(1)}}{f^{(0)}} \right)\left( \frac{\theta}{u}-\frac{2}{u}\right)\nonumber\\
+\frac{1}{2 r_H^{2}f^{(0)}}+\frac{f^{(1)}}{u^{2}f^{2(0)}}\left(\frac{u}{r_H} \right)^{\theta}\mathsf{V}(\varphi)-\frac{f^{(1)}}{4f^{2(0)}}\left(\frac{u}{r_H} \right)^{2z-\theta}\sum_{i}e^{\lambda_{i}\varphi}\mathsf{F}_{iut}^{2}= 0.
\label{E22} 
\end{eqnarray*}

The above set of equations (\ref{E22}) are in general quite difficult to solve for obvious reasons. The solutions corresponding to $ \mathcal{B}^{(1)}(u) $ and $ \mathcal{A}^{(1)}(u) $ could in principle be expressed in terms of the Green's function. However, for the purpose of our present analysis we would be mostly interested to explore the above solutions near the boundary of the space time where the effect of anisotropy gradually diminishes to zero so that the asymptotic structure of the space time remains to that of (\ref{I1}). To do that, let us first consider the equation corresponding to $ \mathcal{B}^{(1)}(u) $ which could be rewritten as,
\begin{eqnarray}
\mathcal{D}_{u}^{(\mathcal{B})}\mathcal{B}^{(1)}(u)=-\frac{e^{\sigma_{\mathcal{B}}}}{2r_H^{2}f^{(0)}(u)}
\end{eqnarray}
where, 
\begin{eqnarray}
\mathcal{D}_{u}^{(\mathcal{B})}&=& \partial_{u}\left( e^{\sigma_{\mathcal{B}}} \partial_u \right)\nonumber\\ 
\sigma_{\mathcal{B}}(\theta,z,u)&=&\left(\frac{\theta -z}{u}-\frac{1}{u}+\frac{f'^{(0)}}{f^{(0)}} \right).
\end{eqnarray}
The corresponding solution could be formally expressed as,
\begin{eqnarray}
\mathcal{B}^{(1)}(u)=\int_{0}^{1} du'\frac{e^{\sigma_{\mathcal{B}}(u')}}{2r_H^{2}f^{(0)}(u')} \mathfrak{G}_{\mathcal{B}}(u,u')
\end{eqnarray}
where, $ \mathfrak{G}_{\mathcal{B}}(u,u') $  is the associated Green's function that satisfies,
\begin{eqnarray}
\mathcal{D}_{u}^{(\mathcal{B})}\mathfrak{G}_{\mathcal{B}}(u,u') = -\delta (u-u').
\end{eqnarray}

Likewise, one could also re-express the equation corresponding to $ \mathcal{A}^{(1)}(u) $ as,
\begin{eqnarray}
\mathcal{D}_{u}^{(\mathcal{A})}\mathcal{A}^{(1)}(u)=-e^{\sigma_{\mathcal{A}}}\Xi^{(\mathcal{A})}(u,\theta , z)
\end{eqnarray}
where,
\begin{eqnarray*}
\mathcal{D}_{u}^{(\mathcal{A})}&=& \partial_{u}\left( e^{\sigma_{\mathcal{A}}} \partial_u \right)\nonumber\\ 
\sigma_{\mathcal{A}}(\theta,z,u)&=&\left(\frac{f'^{(0)}}{f^{(0)}}+\frac{-z+2\theta}{u}-\frac{3}{u} \right)
\end{eqnarray*}
\begin{eqnarray}
\Xi^{(\mathcal{A})}(u, \theta , z)=\frac{f'^{(0)}}{f^{(0)}}\left(\frac{f'^{(1)}}{f'^{(0)}}-\frac{f^{(1)}}{f^{(0)}} \right)\left( \frac{\theta}{u}-\frac{2}{u}\right)+\frac{1}{2 r_H^{2}f^{(0)}}+\frac{f^{(1)}}{u^{2}f^{2(0)}}\left(\frac{u}{r_H} \right)^{\theta}\mathsf{V}(\varphi)\nonumber\\
-\frac{f^{(1)}}{4f^{2(0)}}\left(\frac{u}{r_H} \right)^{2z-\theta}\sum_{i}e^{\lambda_{i}\varphi}\mathsf{F}_{iut}^{2}.
\end{eqnarray}
The corresponding solution could be formally expressed as,
\begin{eqnarray}
\mathcal{A}^{(1)}(u)=\int_{0}^{1} du'e^{\sigma_{\mathcal{A}}(u')}\Xi^{(\mathcal{A})}(u',\theta , z) \mathfrak{G}_{\mathcal{A}}(u,u')
\end{eqnarray}
where, $ \mathfrak{G}_{\mathcal{B}}(u,u') $  is the corresponding Green's function that satisfies,
\begin{eqnarray}
\mathcal{D}_{u}^{(\mathcal{A})}\mathfrak{G}_{\mathcal{A}}(u,u') = -\delta (u-u').
\end{eqnarray}
Note that, here these Green's functions ($  \mathfrak{G}_{\mathcal{A},\mathcal{B}}(u,u') $) are such that they exactly vanish near the effective UV scale of the boundary theory namely, $\mathfrak{G}_{\mathcal{A},\mathcal{B}}(0,u')=0$. As a result, all these effects of anisotropy smoothly go away as we proceed near the boundary of the space time. For the sake of completeness, however, in the following we note down the asymptotic structure corresponding to the functions $ \mathcal{A}^{(1)}(u) $ and $ \mathcal{B}^{(1)}(u) $ namely \footnote{Note that, in order to arrive at the above set of solutions (\ref{E28}), we have imposed the following constraint namely, $ z\geq \frac{2}{2 -\theta} $ where, $ \theta <2 $ for the present example \cite{Alishahiha:2012qu}. A second important point that is to be noted here is that in the asymptotic limit the contribution coming from the anisotropic sector to the matter content of the theory does not play any role in obtaining the solutions. The reason for this lies on the assumption that near the boundary of the space time, the effects of anisotropy are always suppressed compared to that of the leafing order (unperturbed) results as the anisotropy rapidly dies off near the boundary. Therefore, it seems to be only the leading term in (\ref{E21}) that plays the most dominant role while obtaining the solutions near the boundary of the space time.},
\begin{eqnarray}
\mathcal{A}^{(1)}(u)&\approx & \frac{u^2}{4 r_H^2 (2+z-2 \theta )}+\frac{\mathfrak{C}_{a}u^{4+z-2 \theta } }{4+z-2 \theta }\nonumber\\
\mathcal{B}^{(1)}(u)&\approx &\frac{u^2}{4 r_H^2 (z-\theta )}+\frac{\mathfrak{C}_{b}u^{2+z-\theta } }{2+z-\theta }
\label{E28}
\end{eqnarray}
where, $ \mathfrak{C}_{a,b} $ are the integration constants that we fix later on. Since at the end of our calculations, we would be mostly interested to evaluate various observable of the dual gauge theory near its UV cut off ($ \epsilon \sim E_F $), therefore these solutions (\ref{E28}) are extremely important as well as sufficient for the present purpose of the paper.
 
Finally, we re-express the equation corresponding to $ f^{(1)}(u) $ as,
\begin{eqnarray}
f''^{(1)}+\mathfrak{p}(u)f'^{(1)}+\mathfrak{q}(u)f^{(1)}=\mathfrak{g}(u)
\label{E31}
\end{eqnarray}
which is a second order inhomogeneous linear differential equation with coefficients,
\begin{eqnarray*}
\mathfrak{p}(u) = \frac{-3z+2\theta}{u}-\frac{1}{u}-\frac{2\theta +4}{u(1+f^{(0)})}
\end{eqnarray*}
\begin{eqnarray*}
\mathfrak{q}(u) = \frac{(2z-\theta)}{u^{2}(1+f^{(0)})^{2}}(z+(z-\theta)f^{(0)}(2+f^{(0)}))+\frac{\varphi'^{2}}{(1+f^{((0)})^{2}}-\frac{2f^{(0)}(2+\theta (\theta -z))}{u^{2}(1+f^{(0)})^{2}}\nonumber\\
+\frac{2\theta f'^{(0)}}{u(1+f^{(0)})^{2}}-\frac{2(z+1-\theta)f^{(0)}}{u^{2}(1+f^{(0)})^{2}}+\frac{2(2z-\theta)f^{(0)}(2+f^{(0)})}{u^{2}(1+f^{(0)})^{2}}-\frac{4f'^{(0)}}{u(1+f^{(0)})^{2}}\nonumber\\
+\frac{1}{2(1+f^{(0)})^{2}}\left(\frac{u}{r_H} \right)^{2z-\theta } \sum_{i}e^{\lambda_{i}\varphi}\mathsf{F}_{iut}^{2}+\frac{3u^{2}}{2r_H^{\theta}u^{2-\theta}(1+f^{(0)})^{2}}\left( \frac{r_H}{u}\right)^{2(\theta -z)}\sum_{i}e^{\lambda_{i}\varphi}\mathsf{F}_{iut}^{2} 
\end{eqnarray*}
\begin{eqnarray}
\mathfrak{g}(u) =-f'^{(0)}\mathcal{A}'^{(1)}-\frac{4f^{(0)}\mathcal{A}^{'(1)}}{u(1+f^{(0)})}-\frac{(z+1-\theta)f^{(0)}\mathcal{A}'^{(1)}}{u(1+f^{(0)})}+\frac{(2z-\theta)f^{2(0)}\mathcal{A}'^{(1)}}{u(1+f^{(0)})}\nonumber\\
+\frac{2f'^{(0)}\mathcal{A}'^{(1)}}{(1+f^{(0)})}+\frac{1}{r_H^{2}(1+f^{(0)})}.
\end{eqnarray}

The general solution corresponding to (\ref{E31}) could be formally expressed as,
\begin{eqnarray}
f^{(1)}(u)= \mathfrak{y}_{c}+\mathfrak{Y}_{P}
\end{eqnarray}
where, $ \mathfrak{y}_{c} $ is the complementary function that is the solution of the homogeneous equation of the form,
\begin{eqnarray}
f''^{(1)}+\mathfrak{p}(u)f'^{(1)}+\mathfrak{q}(u)f^{(1)}=0
\end{eqnarray}
and $ \mathfrak{Y}_{P} $ is the particular solution that satisfies the inhomogeneous equation itself.

The above equation (\ref{E31}) is not exactly solvable in $ u $ for obvious reasons. However, one could carry out an (approximate) analytic solution near the boundary of the space time which for the present case turns out to be,
\begin{eqnarray}
f^{(1)}(u)\approx \frac{\mathfrak{C}_{f}u^{4+3 z-\theta } }{4+3 z-\theta }+\frac{u^2}{4 r_H^2 (-2-3 z+\theta )} 
\label{E33} 
\end{eqnarray}
subjected to the following constraint namely,
\begin{eqnarray}
(4z-\theta)(4z-\theta +4)-8z^{2}-6z -5=0
\end{eqnarray}
where, $ \mathfrak{C}_{f} $ is some arbitrary integration constant that we fix later on. The other integration constant we set equal to zero in order for the solution to be consistent with the asymptotic boundary condition namely, $ f^{(1)}(u)|_{u \rightarrow 0}\sim 0 $. With the above solutions in hand, we are now in a position to explore the holographic consequences of such anisotropies on the dual field theory near its UV scale. We start with the holographic description of entanglement entropy in the presence of anisotropy.

\section{Entanglement entropy }
The holographic entanglement entropy is defined as the minimal area of the entangling surface (that extends inside the bulk) whose boundary coincides with the boundary of the entangling region namely \cite{Ryu:2006bv}-\cite{Ryu:2006ef},
\begin{eqnarray}
\mathcal{S}_{\mathsf{A}}= \frac{Area(\gamma_{\mathsf{A}})}{4G_N}
\label{E35}
\end{eqnarray}
where, $ \gamma_{\mathsf{A}} $ is the minimal surface that extends in the bulk.

In order to proceed further, we consider a long two dimensional strip as the entangling region at the boundary,
\begin{eqnarray}
-\frac{\mathfrak{l}}{2}\leq x (u) \leq \frac{\mathfrak{l}}{2},~~0 \leq y \leq L
\end{eqnarray}
where, $ \mathfrak{l} $ is the width of the strip. 

The induced metric on the hypersurface turns out to be,
\begin{eqnarray}
ds_{\gamma}^{2}=\left(\frac{u}{r_H} \right)^{\theta -2} \left[ \left( \frac{1}{r_H^{2}f(u)}+e^{\Lambda^{2}(\mathcal{A}^{(1)}+\mathcal{B}^{(1)})}x'^{2} \right)du^{2} +e^{\Lambda^{2}(\mathcal{A}^{(1)}-\mathcal{B}^{(1)})} dy^{2}\right]
\label{E36} 
\end{eqnarray}
where, prime denotes derivative w.r.t. the radial coordinate $ u $.

With the above choice of the metric (\ref{E36}), the corresponding area functional turns out to be,
\begin{eqnarray}
\mathsf{A}=L \int du~ u^{\theta -2} \sqrt{e^{\Lambda^{2}\mathcal{A}^{(1)}}\left( \frac{e^{-\Lambda^{2}\mathcal{B}^{(1)}}}{f(u)}+e^{\Lambda^{2}\mathcal{A}^{(1)}}x'^{2}\right) }.
\label{E37}
\end{eqnarray}

From the structure of (\ref{E37}), it is indeed clear that the momentum conjugate to $ x $ is conserved namely,
\begin{eqnarray}
\frac{e^{2\Lambda^{2}\mathcal{A}^{(1)}(u)}x'(u)}{\sqrt{e^{\Lambda^{2}\mathcal{A}^{(1)}(u)}\left( \frac{e^{-\Lambda^{2}\mathcal{B}^{(1)}(u)}}{f(u)}+e^{\Lambda^{2}\mathcal{A}^{(1)}(u)}x'^{2}\right) }}=\left(\frac{u_T}{u} \right)^{\theta -2} e^{\Lambda^{2}\mathcal{A}^{(1)}(u_T)}
\label{E38}
\end{eqnarray}
where, $ u_T $ corresponds to the turning point on the hypersurface such that, $ x' \rightarrow \infty $ as $ u \rightarrow u_T $.

Using (\ref{E38}), one could immediately express the width of the strip as,
\begin{eqnarray}
\mathfrak{l}= 2u_T \int_{0}^{1} d\vartheta~ \mathfrak{K}(\vartheta , \Lambda)
\label{E40}
\end{eqnarray}
where, $ \vartheta = \frac{u}{u_T} $ is a new integration variable. The integrand $ \mathfrak{K}(\vartheta , \Lambda) $ could be formally expressed as,
\begin{eqnarray}
\mathfrak{K}(\vartheta , \Lambda) &=& \frac{\vartheta^{(2-\theta)}f^{-1/2}(\vartheta)e^{\frac{\Lambda^{2}}{2}(2\mathcal{A}^{(1)}(u_T)+\mathcal{A}^{(1)}(\vartheta)-\mathcal{B}^{(1)}(\vartheta))}}{\sqrt{e^{4\Lambda^{2}\mathcal{A}^{(1)}(\vartheta)}-\vartheta^{2(2-\theta)}e^{2\Lambda^{2}(\mathcal{A}^{(1)}(u_T)+\mathcal{A}^{(1)}(\vartheta))}}}\nonumber\\
&=&\mathfrak{K}^{(0)}(\vartheta)\left(1+\Lambda^{2}\mathfrak{M}(\vartheta) \right) +\mathcal{O}(\Lambda^{4})
\end{eqnarray}
where,
\begin{eqnarray}
\mathfrak{M}(\vartheta)&=&\frac{\mathcal{A}^{(1)}(u_T)-\mathcal{A}^{(1)}(\vartheta)}{(1-\vartheta^{2(2-\theta)})}-\frac{1}{2}\left(\mathcal{A}^{(1)}(\vartheta)+\mathcal{B}^{(1)}(\vartheta)+\frac{f^{(1)}}{f^{(0)}} \right)\nonumber\\
\mathfrak{K}^{(0)}(\vartheta)&=&\frac{\vartheta^{(2-\theta)}}{\sqrt{f^{(0)}(1-\vartheta^{2(2-\theta)})}} 
\end{eqnarray}

In order to evaluate the above integral (\ref{E40}), we assume that the width of the strip is quite small such that, $ u_T \ll 1 $. This assumption could eventually be translated into the fact that we are probing our system in the low temperature regime where we assume that the entangling surface does not extend much deeper into the bulk space time \cite{Dong:2012se} . 

Finally, using (\ref{E28}) and (\ref{E33}) and setting, $ z=2, \theta =d-1=1 $ we obtain,
\begin{eqnarray}
\mathfrak{l}\approx 2u_T \left( 1+\Lambda^{2} \mathfrak{a}\right) 
\label{E43}
\end{eqnarray}
where,
\begin{eqnarray}
\mathfrak{a}=\frac{7}{5}-\frac{159 \pi }{1024}.
\end{eqnarray}

With (\ref{E43}) in hand, our final task would be to compute the area functional (\ref{E37}) in order to compute the holographic entanglement entropy (\ref{E35}) in the presence of the anisotropy. We first note down the area functional (\ref{E37}) which for the present case turns out to be,
\begin{eqnarray}
\mathsf{A}=L\int_{\epsilon/u_T}^{1}\frac{d\vartheta}{\vartheta}\sqrt{e^{\Lambda^{2}\mathcal{A}^{(1)}(\vartheta)}\left( \frac{e^{-\Lambda^{2}\mathcal{B}^{(1)}(\vartheta)}}{f(\vartheta)}+e^{\Lambda^{2}\mathcal{A}^{(1)}(\vartheta)}\mathfrak{K}^{2}(\vartheta , \Lambda)\right) }
\label{E45}
\end{eqnarray}
where, $ \epsilon (\sim E_{F}) $ is the so called \textit{effective} UV cut-off that we are finally interested in. 

Expanding the integrand (\ref{E45}) upto leading order in the anisotropy we obtain,
\begin{eqnarray}
\mathsf{A}=L\int_{\epsilon/u_T}^{1}\frac{d\vartheta}{\vartheta\sqrt{f^{(0)}(1-\vartheta^{2})}}\left(1+\frac{\Lambda^{2}}{2}\left( 2\vartheta^{2}-\mathcal{B}^{(1)}+\frac{(1-2\vartheta^{4}-3\vartheta^{2})\mathcal{A}^{(1)}}{2(1-\vartheta^{2})}-\frac{f^{(1)}}{f^{(0)}}\right)\right)+\mathcal{O}(\Lambda^{4}).
\label{E46}
\end{eqnarray}

Clearly, the above area functional (\ref{E46}) is the generalization of the previous analysis \cite{Alishahiha:2012qu} in the presence of the anisotropy. In our analysis we are interested to explore the behaviour of the above area functional (\ref{E46}) near its UV cut-off scale ($ \epsilon $) which finally yields,
\begin{eqnarray}
\mathsf{A}\equiv L\log\left( \frac{\mathfrak{l}}{\epsilon}\right)+L \mathcal{K}(\Lambda^{2},\epsilon , \mathfrak{l})+\mathsf{F}(\Lambda^{2},\mathsf{M},\mathsf{Q}^{2},\epsilon , \mathfrak{l}) 
\label{E47}
\end{eqnarray}
where,
\begin{eqnarray}
\mathcal{K}(\Lambda^{2},\epsilon , \mathfrak{l})=-\Lambda^{2}\mathfrak{a}+\log 4 -\frac{\epsilon^{2}}{\mathfrak{l}^{2}}(1+2\Lambda^{2}\mathfrak{a}^{2})+\Lambda^{2}\left(1-\frac{2\epsilon^{2}}{\mathfrak{l}^{2}}+\frac{31}{10}\right)\nonumber\\
+ \frac{\Lambda^{2} T}{64 (4\mathsf{Q}^{2}-3\mathsf{M})}\left(\frac{17 \epsilon^{2}}{7}+\frac{145\mathfrak{l}^{2}}{42}\right) +\mathcal{O}(\Lambda^{4}\epsilon^{3}) 
\end{eqnarray}
where, $ T $ is temperature of the black brane. Note that, the other function $\mathsf{F}(\Lambda^{2},\mathsf{M},\mathsf{Q}^{2},\epsilon , \mathfrak{l})  $ also contains some further temperature corrections that includes contributions from the anisotropic sector of the theory. However, in the low temperature regime of the theory ($ u \ll 1 $), these effects are highly suppressed compared to those of the leading order terms in (\ref{E47}).

Substituting, (\ref{E47}) into (\ref{E35}), we finally obtain the entanglement entropy as,
\begin{eqnarray}
\mathcal{S}_{\mathsf{A}}= \frac{L}{4G_N }\left( \log\left( \frac{\mathfrak{l}}{\epsilon}\right)+ \mathcal{K}(\Lambda^{2},\epsilon ,\mathfrak{l})\right) +..~..
\label{E48}
\end{eqnarray}
The appearance of the logarithmic divergence above in (\ref{E48}) is an artefact of the presence of $ \mathcal{O}(N^{2}) $ Fermi surface in the dual field theory \cite{Dong:2012se}-\cite{Alishahiha:2012qu}. However, apart from having this usual logarithmic divergence, in the present scenario we also encounter additional interesting contributions to the holographic entanglement entropy due to the presence of the anisotropy in the bulk description. This anisotropy is associated with the breaking of the rotational invariance of the (hidden) Fermi surfaces of the dual QFT. Now it turns out that one could clearly divide the full contribution, $\mathcal{K}(\Lambda^{2},\epsilon ,\mathfrak{l})) $ into two pieces, one that does not depend on the temperature of the system and the other part that explicitly depends on the temperature of the thermal bath. The part that does not depend on the temperature ($ T $), might be regarded as the anisotropic contribution to the entanglement entropy associated with the compressible quark matter in its ground state. On the other hand, the finite temperature piece in the entanglement entropy corresponds to the anisotropic contribution associated with the raising of the compressible state of fractionalized fermionic excitation at non zero temperature. 
\section{Wilson loop}
In this section, based on the AdS/CFT prescription, our goal is to compute the Wilson loop ($ \mathit{W} $) for strongly coupled field theories in the presence of the hyperscaling violation near the UV scale of the theory. In the language of the Gauge/gravity duality, the Wilson loop could be thought of as represented by the the end points of an open string whose world sheet has the boundary on the $ D $ brane that is pushed towards the boundary of the space time under consideration. In the t'Hooft limit, the holographic prescription for evaluating the Wilson loop turns out to be \cite{Maldacena:1998im},
\begin{eqnarray}
\langle \mathit{W}\rangle =e^{-S_{NG}}
\end{eqnarray}
where, $ S_{NG} $ stands for the on shell extremal Nambu-Goto (NG) action for the string world sheet,
\begin{eqnarray}
S_{NG}=\frac{1}{2 \pi \alpha'}\int d^{2}\sigma \sqrt{-\zeta}.
\end{eqnarray}

In our analysis, we parametrize string embedding by the function, $ u=u(x) $ and denote derivatives as, $ u'(x) $. With these assumptions in hand, the induced metric on the string world sheet turns out to be,
\begin{eqnarray}
ds^{2}_{NG}=\left(\frac{r_H}{u} \right)^{-\theta} \left[-\left( \frac{r_H}{u}\right)^{2z}f(u)dt^{2}+\left( \frac{u'^{2}}{u^{2}f(u)}+e^{\mathcal{A}(u)+\mathcal{B}(u)}\right) dx^{2}\right]
\end{eqnarray} 
which finally yields the NG action as\footnote{Where, we have set, $ z=2 $ and $ \theta =1 $.},
\begin{eqnarray}
S_{NG}=\frac{\tau}{2 \pi \alpha'}\int ~dx~ \left( \frac{r_H}{u^{2}}\right) \sqrt{u'^{2}+r_H^{2}f(u)e^{\Lambda^{2}(\mathcal{A}^{(1)}(u)+\mathcal{B}^{(1)}(u))}} 
\label{E52} 
\end{eqnarray}
where, we have considered world sheet coordinates as, $ \sigma^{0}=t $ and $ \sigma^{1}=x $. Furthermore, we have replaced the time integral as, $ \tau = \int dt $.

The first integral of the equation of motion that readily follows from (\ref{E52}) could be formally expressed as,
\begin{eqnarray}
\frac{u^{2}\sqrt{u'^{2}+r_H^{2}f(u)e^{\Lambda^{2}(\mathcal{A}^{(1)}(u)+\mathcal{B}^{(1)}(u))}}}{r_H^{3} f(u)e^{\Lambda^{2}(\mathcal{A}^{(1)}(u)+\mathcal{B}^{(1)}(u))}}=constant.
\label{E53}
\end{eqnarray}

In order to set the constant appearing on the r.h.s. of (\ref{E53}), we first note that the boundary conditions for the function $ u(x) $ turns out to be,
\begin{eqnarray}
u (x=-\mathfrak{d}/2)=0=u(x=\mathfrak{d}/2)
\end{eqnarray} 
where, $ \mathfrak{d} $ is the separation between the quark ($ \mathfrak{q} $) and the anti quark ($ \bar{\mathfrak{q}} $) pair. By the symmetry of the problem, therefore, clearly there is value for $ x(=0) $ where $ u $ is maximal (say, $ u_{\ast} $). Which eventually implies that at this point, $ u'(x=0)=0 $. Considering all these facts, from (\ref{E53}) we note that,
\begin{eqnarray}
u' = \pm r_H \left( \frac{u_{\ast}}{u}\right)^{2} \sqrt{\frac{f(u)e^{\Lambda^{2}(\mathcal{A}^{(1)}(u)+\mathcal{B}^{(1)}(u))}}{f(u_{\ast})e^{\Lambda^{2}(\mathcal{A}^{(1)}(u_{\ast})+\mathcal{B}^{(1)}(u_{\ast}))}}}\times \nonumber\\
\left(f(u)e^{\Lambda^{2}(\mathcal{A}^{(1)}(u)+\mathcal{B}^{(1)}(u))}-\left(\frac{u}{u_{\ast}} \right)^{4} f(u_{\ast})e^{\Lambda^{2}(\mathcal{A}^{(1)}(u_{\ast})+\mathcal{B}^{(1)}(u_{\ast}))}\right)^{1/2}.
\label{E55} 
\end{eqnarray}

Eq.(\ref{E55}) could be inverted in order to obtain,
\begin{eqnarray}
\mathfrak{d}=\frac{2u_{\ast}}{r_H}\int_{0}^{1}\frac{\xi^{2}d\xi}{\mathcal{F}(\xi , \Lambda^{2})}
\label{E56}
\end{eqnarray}
where, we have defined a new variable, $ \xi =\frac{u}{u_{\ast}} $ and,
\begin{eqnarray}
\mathcal{F}(\xi , \Lambda^{2})=\sqrt{\frac{f(\xi)e^{\Lambda^{2}(\mathcal{A}^{(1)}(\xi)+\mathcal{B}^{(1)}(\xi))}}{f(u_{\ast})e^{\Lambda^{2}(\mathcal{A}^{(1)}(u_{\ast})+\mathcal{B}^{(1)}(u_{\ast}))}}}\times \nonumber\\
\left(f(\xi)e^{\Lambda^{2}(\mathcal{A}^{(1)}(\xi)+\mathcal{B}^{(1)}(\xi))}-\xi^{4} f(u_{\ast})e^{\Lambda^{2}(\mathcal{A}^{(1)}(u_{\ast})+\mathcal{B}^{(1)}(u_{\ast}))}\right)^{1/2}.
\end{eqnarray}

The above integral (\ref{E56}), is in general quite difficult to evaluate analytically. However, under certain specific assumption namely, $ u_{\ast}\ll 1 $ this could performed analytically which finally yields, 
\begin{eqnarray}
\mathfrak{d}=\frac{2\sqrt{\pi }u_{\ast}}{r_H}\frac{ \Gamma\left[\frac{3}{4}\right]}{\Gamma\left[\frac{1}{4}\right]}\left(1+\frac{\Lambda^{2}}{2\sqrt{\pi}}\frac{ \Gamma\left[\frac{1}{4}\right]}{\Gamma\left[\frac{3}{4}\right]}\left( \frac{16}{15}+\frac{\pi }{8}-\frac{3 \sqrt{\pi } \Gamma\left[\frac{3}{4}\right]}{16 \Gamma\left[\frac{9}{4}\right]}\right)  \right) +\mathcal{O}(\Lambda^{4}).
\end{eqnarray}

We now focus towards the computation of the on-shell action (\ref{E52}) for the string which will finally enable us to determine the quark-antiquark potential for the boundary theory. The on-shell action (\ref{E52}) turns out to be,
\begin{eqnarray}
S_{NG}=\frac{\tau r_H}{ \pi \alpha' u_{\ast}}\int_{\epsilon/u_{\ast}}^{1} ~\frac{d\xi}{\xi^{2}}~  \sqrt{1+\mathcal{H}(\xi , \Lambda^{2})}
\label{E59}  
\end{eqnarray}
where, $ \epsilon $ is the so called UV cut off scale for the theory and the function $ \mathcal{H}(\xi , \Lambda^{2}) $ could be formally expressed as ,
\begin{eqnarray}
\mathcal{H}(\xi , \Lambda^{2}) = \frac{\xi^{4}}{1-\xi^{4}}\left(1-\Lambda^{2} \frac{(\xi^{9}+\xi^{4}+\xi^{3}-3)}{(1-\xi^{4})}\right) +\mathcal{O}(\Lambda^{4}).
\label{E60}
\end{eqnarray}

Using (\ref{E60}), the on-shell action (\ref{E59}) finally turns out to be,
\begin{eqnarray}
S_{NG}=\tau \mathsf{V}_{\mathfrak{q}\bar{\mathfrak{q}}}
\end{eqnarray}
where,
\begin{eqnarray}
| \mathsf{V}_{\mathfrak{q}\bar{\mathfrak{q}}}|&=&\frac{2\sqrt{\lambda}}{\sqrt{\pi}\mathfrak{d}}\frac{ \Gamma\left[\frac{3}{4}\right]}{\Gamma\left[\frac{1}{4}\right]}\left(|K[-1]-E[-1]|-\frac{\Lambda^{2}}{2} \mathcal{Z} \right)+\mathcal{O}(1/\epsilon)\nonumber\\
&=& \mathsf{V}^{(0)}_{\mathfrak{q}\bar{\mathfrak{q}}}-\Lambda^{2}\mathsf{V}^{(1)}_{\mathfrak{q}\bar{\mathfrak{q}}}+\mathcal{O}(1/\epsilon)
\label{E63}
\end{eqnarray}
is the potential energy between the $  \mathfrak{q}\bar{\mathfrak{q}}$ separated at a distance $ \mathfrak{d} $ from each other and $ \alpha \sim \sqrt{\lambda} $ is the famous t'Hooft coupling for the boundary field theory. Here, $ K $ and $ E $ are respectively the Elliptic integrals of first kind and second kind. The function $ \mathcal{Z} $ could be formally expressed as,
\begin{eqnarray}
\mathcal{Z}=\frac{1}{\sqrt{\pi}}\frac{ \Gamma\left[\frac{1}{4}\right]}{\Gamma\left[\frac{3}{4}\right]}\left( \frac{16}{15}+\frac{\pi }{8}-\frac{3 \sqrt{\pi } \Gamma\left[\frac{3}{4}\right]}{16 \Gamma\left[\frac{9}{4}\right]}\right) (K[-1]-E[-1])+\left( \frac{4}{3}+\frac{\pi }{4}\right). 
\end{eqnarray}

Before we conclude, it is customary to make a few comments about the above result in (\ref{E63}). First of all, the $ 1/\mathfrak{d} $ dependence in our result reminds us about the famous Coulomb interaction between the $  \mathfrak{q}\bar{\mathfrak{q}} $ pair separated at a distance $ \mathfrak{d} $ from each other. The $ \sqrt{\lambda} $ dependence is the consequence of the strong coupling of the theory. This result also reminds us about the zero temperature results of the theory and which is perfectly consistent with the fact that we have confined ourselves in the regime near the boundary of the space time namely, $ u_{\ast}\ll 1 $. The $ 1/\epsilon $ divergence corresponds to the usual UV divergence in the static limit \cite{Maldacena:1998im}. 

Let us now turn over towards the anisotropic contributions of the theory. From the exact numerical values of the corresponding functions the value of $ \mathcal{Z} $ turns out to be,
\begin{eqnarray}
\mathcal{Z} \approx 1.02332
\label{ee70} 
\end{eqnarray}
which therefore clearly reduces and/or suppresses the Coulomb potential between the $ \mathfrak{q}\bar{\mathfrak{q}} $ pair.  

One could view (\ref{E63}) as being the potential energy between the fractionalized gauge charged (\textit{quark} like) excitation of the theory that is distributed over the hidden Fermi surfaces of the boundary as mentioned in the introductory note.  The Coulomb like nature of the above potential (\ref{E63}) further confirms the fact that the theory does not confine near its effective dynamical scale and this is consistent with the fact that the gauge fields do not confine near the effective dynamical scale of the theory. From the above analysis, it further turns out that once we deform (breaking the rotational invariance of Fermi surfaces) the theory by means of the anisotropy, then the potential energy gets further reduced due to the presence of the factor $ \mathcal{Z} $ (\ref{ee70}) which therefore acts as a physical screening between the fermions.

\section{Summary and final remarks}
Let us now summarize the key findings of our analysis.  The primary motivation of the present analysis was to study the anisotropic Fermi surfaces under a holographic set up, where the anisotropy in the bulk is induced due to some mass less scalar excitation that explicitly break the rotational invariance of the original gravitational solution. When we probe our system through non local observable like, the entanglement entropy it turns out that it receives both thermal as well as non thermal contributions due to anisotropy present in the bulk description. We also explore Wilson operator for such systems and it turns out that the anisotropy suppresses the potential energy between the quark anti- quark pair at the boundary. In summary, we hope that, in the near future, the present analysis of this paper would play an important role in order to understand various time dependent phenomena for a class of hyperscaling violating QFTs at strong coupling.
\\ \\
{\bf {Acknowledgements :}}
 The author would like to acknowledge the financial support from UGC (Project No UGC/PHY/2014236).\\ \\
 \textbf{Appendix I}\\
 Equation for $ f $:
 \begin{eqnarray*}
f'' +\frac{f'}{u}(1-3z+2\theta+u\mathcal{A}')+\frac{f(2z-\theta)(f(z-\theta)+z)}{u^{2}(f+1)}-\frac{f(\mathcal{A}'^{2}-\mathcal{B}'^{2})}{(f+1)}-\frac{2(\theta f (\theta -z)+u \theta f')}{u^{2}(f+1)}\nonumber\\
+\frac{\mathcal{A}'}{u^{2}(f+1)}\left(uf(2+\theta)-uf^{2}(2z-\theta)-2u^{2}f\left(\frac{f'}{f}+\frac{1-z+2\theta}{u} \right)  \right)-\Lambda^{2}\frac{e^{-(\mathcal{A}+\mathcal{B})}}{u^{2}(f+1)}+\frac{2\mathsf{V}(\varphi)}{u^{2}(f+1)}\left(\frac{u}{r_H} \right)^{\theta}\nonumber\\
+\frac{1}{2u^{2}(f+1)}\left(\frac{u}{r_H} \right)^{\theta}\sum_{i}e^{\lambda_{i}\varphi}\mathsf{F}_{i}^{2} +\frac{2}{(f+1)}\frac{1}{ r_H^{\theta}u^{2-\theta}}\left(T^{t}\ _{t}+T^{u}\ _{u}\right)=0.
\end{eqnarray*}
\textbf{Appendix II}\\
 Equation for $ f^{(1)} $:
\begin{eqnarray*}
f^{''(1)}+\left( \frac{-3z+2\theta}{u}-\frac{1}{u}\right) f'^{(1)}+f'^{(0)}\mathcal{A}'^{(1)}+\frac{(2z-\theta)f^{(1)}}{u^{2}(1+f^{(0)})^{2}}(z+(z-\theta)f^{(0)}(2+f^{(0)}))\nonumber\\
+\frac{\varphi'^{2}f^{(1)}}{(1+f^{((0)})^{2}}-\frac{2f^{(0)}f^{(1)}(2+\theta (\theta -z))}{u^{2}(1+f^{(0)})^{2}}+\frac{4f^{(0)}\mathcal{A}^{'(1)}}{u(1+f^{(0)})}-\frac{2\theta f'^{(0)}}{u(1+f^{(0)})}\left( \frac{f'^{(1)}}{f'^{(0)}}-\frac{f^{(1)}}{1+f^{(0)}}\right)\nonumber\\
-\frac{2(z+1-\theta)f^{(0)}}{u^{2}(1+f^{(0)})} \left(-\frac{u}{2}\mathcal{A}'^{(1)}+\frac{f^{(1)}}{f^{(0)}(1+f^{(0)})} \right)+\frac{2(2z-\theta)f^{2(0)}}{u^{2}(1+f^{(0)})} \left(-\frac{u}{2}\mathcal{A}'^{(1)}+\frac{f^{(1)}(2+f^{(0)})}{f^{(0)}(1+f^{(0)})} \right)\nonumber\\
+\frac{4f'^{(0)}}{u(1+f^{(0)})}\left(-\frac{u}{2}\mathcal{A}'^{(1)}+ \frac{f'^{(1)}}{f'^{(0)}}-\frac{f^{(1)}}{1+f^{(0)}}\right)-\frac{1}{r_H^{2}(1+f^{(0)})} -\frac{2\mathsf{V}(\varphi)f^{(1)}}{u^{2}(1+f^{(0)})^{2}}\left(\frac{u}{r_H} \right)^{\theta}\nonumber\\
+\frac{f^{(1)}}{2(1+f^{(0)})^{2}}\left(\frac{u}{r_H} \right)^{2z-\theta } \sum_{i}e^{\lambda_{i}\varphi}\mathsf{F}_{iut}^{2}+\frac{2f^{(1)}}{r_H^{\theta}u^{2-\theta}(1+f^{(0)})^{2}}\left( \mathsf{V}(\varphi)+\frac{3u^{2}}{4}\left( \frac{r_H}{u}\right)^{2(\theta -z)}\sum_{i}e^{\lambda_{i}\varphi}\mathsf{F}_{iut}^{2} \right) =0.
\end{eqnarray*}


\end{document}